\documentclass[twocolumn,pra,showpacs,preprintnumbers,amsmath,amssymb]{revtex4}
\usepackage{epsfig,amsmath}

\usepackage{graphicx}% Include figure files
\usepackage{dcolumn}% Align table columns on decimal point

\newcommand{\beq}{\begin{equation}}
\newcommand{\eeq}{\end{equation}}
\newcommand{\beqa}{\begin{eqnarray}}
\newcommand{\eeqa}{\end{eqnarray}}
\newcommand{\lam}{\lambda}
 \newcommand{\rh}{\rho}
\newcommand{\ga}{\gamma} 
 
 \newcommand{\si}{\sigma}
 \newcommand{\om}{\omega}
 
\newcommand{\ra}{\rangle}

\def\pra#1{{ Phys.\ Rev. A\/} {\bf#1}} \def\prb#1{{ Phys.\ Rev. B\/} {\bf#1}}
 \def\prl#1{{ Phys.\ Rev.\
Lett.} {\bf#1}}

\begin{document}

%\date{4 OCT  2005}

%for non-revtex

\title{Negative Entanglement Measure, and What It Implies}
\author{Ting  Yu}
\email{ting@pas.rochester.edu}

   %\altaffiliation{}
   %Lines break automatically or can be f%orced with \\

\author{J.\ H.\ Eberly}
\email{eberly@pas.rochester.edu}

%\author{}
%\author{}
\affiliation{Rochester Theory Center for Optical
Science and Engineering and the Department of Physics and Astronomy
University of Rochester, Rochester, New York 14627}

\begin{abstract}
In this paper, we extract from concurrence its variable part, denoted $\Lambda$, and use $\Lambda$ as a time-dependent measure of ``distance", either postive or negative, from the separability boundary. We use it to investigate entanglement dynamics of two isolated but initially entangled qubits, each coupled to  its own environment. 
\vspace*{0.5cm}

\end{abstract}


%turn off for non-revetex

\maketitle  %turn off for nonrevtex
%this is for revtexprotocols
%\newpage

%%%%%%%%%%%%%%%%%%%%%%%%%%%%%%%%%%%%%%%%%%%%
\section{Introduction}
%%%%%%%%%%%%%%%%%%%%%%%%%%%%%%%%%%%%%%%%%

Entanglement is arguably the most  intriguing feature of  quantum 
mechanics \cite{EPRpaper}.  In a ideal world, the entanglement needed 
for quantum information processing would be stable and uncorrupted. 
But in reality, the interaction of a quantum system with its 
surroundings is unavoidable because no real physical system can be 
isolated completely from  external noise. The effect of noise on {\em pairs of quantum systems} is particularly interesting when the systems are relatively well isolated from noise and undergo relaxation individually rather slowly, but may be vulnerable to qualitatively different relaxation channels in regard to mutual coheence, i.e., to entanglement. Five years ago such a difference was pointed out by the group of Knight \cite{PLKetal02}. The present paper can be considered an indirect extension of the results reported there.

More recently new features have been discovered for entanglement decoherence as a time-dependent process. It has begun to be the subject of wide interest  
\cite{Yu-Eberly02,Yu-Eberly03,Yu-Eberly04,diosi,Halliwell04l,Jakobczyk-Jamroz04,YuEberly-arX05,privman,huetal,phys,Santos-etal06,Ban06,Yu-Eberly06d,Yonac-Yu-Eberly060c,Ficek-Tanas06}. 
Contrary to intuition based on experience with single qubit 
decoherence, we have shown \cite{Yu-Eberly04} that entanglement may 
decay to zero in a finite time (and remain zero for at least a finite 
time).

\begin{figure}[h]
%%\begin{figure}[!b]
\includegraphics[width=5 cm]{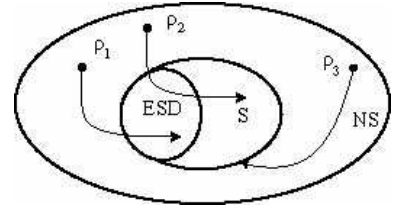}
%\epsfig{file=FigW-phs.eps, width=6.0 cm}
% Here is how to import EPS art
\caption{\footnotesize \label{solutionspace} Quantum trajectories in a non-metric ``evolution space" of density matrices of two qubits.  The space of the two qubits can be divided into the region of separable states denoted by $S$, and entangled states denoted by $NS$.  Within the separable states $S$ we identify an important subset (ESD) that includes all the states that reach separability abruptly in a finite time rather than asymptotically.  The diagram shows the three initially entangled states $\rho_1,\rho_2$ and $\rho_3$ evolving into three separable states.} 
\end{figure}

This finite-time decay is  called entanglement sudden death (ESD). It 
is a feature of nonlocal quantum coherence, one that introduces a 
new bipartite element into quantum open system behavior 
\cite{Yu-Eberly06d,Yu-Eberly06prl}. ESD adds an element of complexity to the commonly encountered state-space diagrams showing evolution along various routes that, in the normal cases of relaxation lead to separability. An example is shown in Fig. \ref{solutionspace}, where trajectories all begin as entangled states (in the non-separable space) and evolve toward separability. 

Here we identify the boundary between the $S$ and $NS$ regions as a border that establishes the origin of a generalized ``coordinate", which we define.  In this context we examine the evolution of different initial $NS$ states under different decoherence mechanisms as their trajectories pass the boundary, i.e., as their coordinate distance from the boundary passes from negative to positive values. We show situations in which this border-crossing behavior is (a) monotonic or (b) periodically repeating, or (c) what we think of as the most familiar type of decoherence evolution,  reaching the border but not crossng it.

%%%%%%%%%%%%%%%%%%%%%%%%%%%%%%%%%%%%%%%%%%%%
\section{The $\Lambda$ Distance for Pure and Mixed States}
%%%%%%%%%%%%%%%%%%%%%%%%%%%%%%%%%%%%%%%%%%%%

For a two-qubit state described by density matrix $\rho$, Wootters' concurrence $C(\rho)$ is the most-used  measure of entanglement (non-separability) because of its relative ease of direct calculation \cite{Wootters}, and its conveniently normalized range from $C=0$ for a separable state to $C=1$ for a maximally entangled state. It should be emphasized that the boundary of separability of a state is independent of the entanglement measure, and in this respect concurrence is exactly in accord with entropy of formation, negativity and tangle, while many other familiar coherence concepts such as purity or non-zero diagonal matrix elements have no reliable relation to entanglement.

For two qubits, the concurrence may be calculated explicitly from the density matrix $\rho$ for qubits A and B:
\beq
C(\rh)=\max\left(0,\sqrt{\lam_1}-\sqrt{\lam_2}-\sqrt{\lam_3}-\sqrt{\lam_4}\,\,\right),
\eeq
where the quantities $\lam_i$ are the eigenvalues in decreasing order 
of the matrix
\beq \zeta = \rho(\sigma^A_y\otimes
\sigma^B_y)\rho^*(\sigma^A_y\otimes
\sigma^B_y),\label{concurrence}
\eeq
where $\rh^*$ denotes the
complex conjugation of $\rh$ in the standard basis $|+,+\rangle, 
|+,-\rangle, |-,+\rangle, |-,-\rangle$ and $\si_y$ is the Pauli 
matrix expressed in the same basis as:

\beq \si^{A,B}_y =
\begin{pmatrix}
0 \,\,& \,\,\,-i \\
i \,\,& \,\,\,\, 0
\end{pmatrix}.
\eeq
The relation among these eigenvalues is of course the essence of the 
concurrence, so for convenience we define the quantity $\Lambda$:
\beq
\Lambda \equiv \sqrt{\lam_1}-\sqrt{\lam_2}-\sqrt{\lam_3}-\sqrt{\lam_4}.
\eeq
It is obvious that $\Lambda$ can serve as a one-dimensional quantum measure of the ``distance" of a state away from the boundary and whether it is on the positive (entangled) or negative (disentangled) side of the solution space of Fig. \ref{solutionspace}. 

A focus on  $\Lambda$ is prompted because it contains different information than the concurrence 
$C(\rho)$ from which it originates. To see this and to begin our  
discussion, we evaluate $\Lambda$ for an arbitrary pure state that is 
separable (not entangled): $\rho^{AB} = |\Psi^{AB}\rangle \langle 
\Psi^{AB}|$,
where
$$ |\Psi^{AB}\ra = \Big( a|+\rangle_A + b|-\rangle_A \Big) \otimes 
\Big( c|+\rangle_B + d|-\rangle_B \Big).$$
It is easy to check that both $C(\rho) = 0$ and the distance $\Lambda 
= 0$ for this case.

From these comments we show that distinctions between $C$ and $\Lambda$ make $\Lambda$  more useful. For example, we see that, given any density matrix $\rho$, if $\Lambda(\rho)$ is strictly negative, $\Lambda < 0$ rather than $\Lambda \le 0$, it implies that $\rho$ must be {\em both} mixed and separable.  Note that this is more than the value of $C(\rho)$ provides. It cannot tell if the state $\rho$ is mixed or not. Another unusual property of $\Lambda$ during decoherence evolution is that its final value will generally depend on the system's initial state, even under fully disentangling evolution, whereas a fully disentangled concurrence must have the value 0 no matter what the initial state was. From the final value of $\Lambda$ one can determine if ESD has occurred, but not from the final value of concurrence.

Clearly it is  important to be aware if a boundary-crossing transition occurs for a bipartite $A-B$ unit  in an active quantum network; and the existence of entanglement sudden death (ESD) signals that the transition  may occur abruptly and in that sense very early, as experimental work is beginning to demonstrate \cite{RioGroup}. Here we describe the way ESD relates to changes in the $\Lambda$ distance for several state trajectories. In all cases we will suppose that $A$ and $B$ are initially entangled and experience relaxation that drives the pair to a separable state.

%%%%%%%%%%%%%%%%%%%%%%%%%%%%%%%%%%%%%%%%%%%%
\section{Phase-relaxing evolution of $\Lambda$}
%%%%%%%%%%%%%%%%%%%%%%%%%%%%%%%%%%%%%%%%%%%%
\label{phase-evolution}

First we consider the time evolution of the $\Lambda$ measure under  broadband dephasing. The following familiar spin Hamiltonian \cite{Yu-Eberly02,Yu-Eberly03,Yu-Eberly06d} provides a good illustration. We take  $\hbar=1$ and write:
\beq
\label{hamiltonian}H(t)=-\frac{1}{2}\mu \Big(b_A(t)\si_z^A
+b_B(t)\si_z^B \Big),
\eeq
where $\mu$  stands for a common gyromagnetic ratio and the magnetic fields $b_A(t)$ and $b_B(t)$ are assumed to be weak, independent, statistically stationary and noisy. We assume their bandwidths are sufficiently smooth and broadband in nature to exclude the existence of decoherence-free subspace effects. They have the average values:
\beqa
<b_i(t)> &=& 0, \\
< b_i(t)b_j(t')> &=& \frac{\Gamma_i}{\mu^2}\delta_{ij}\delta(t-t'),\,\,
i,j=A,B,\label{cor2a}
\eeqa
where the $\Gamma_{i}$ are the phase damping rates of qubits $A$ and 
$B$. For this interaction, the time dependence of density 
matrix $\rho^{AB}(t)$ of the two qubits has been found \cite{Yu-Eberly03} and is given by:
\beqa\label{ch}
&& \rh^{AB}(t)=\nonumber\\
&& = \begin{pmatrix}
\rho_{11} &\ga_B\rho_{12} &  \ga_A\rho_{13} & \ga_A\ga_B\rho_{14}\\
\ga_B\rho_{21} &\rho_{22}&  \ga_A\ga_B\rho_{23}& \ga_A\rho_{24}\\
\ga_A\rho_{31}&\ga_A\ga_B\rho_{32}&\rho_{33}& \ga_B\rho_{34}\\
\ga_A\ga_B\rho_{41}&\ga_A\rho_{42}&\ga_B\rho_{43}& \rho_{44}
\end{pmatrix},
\eeqa
where the $\rho_{ij}$ elements here are initial values and the 
$\gamma$'s contain the time dependences:
\beq
\ga_A={e}^{-{t}/{2T^A_2}},\,\,\,\ga_B={e}^{-{t}/{2T^B_2}}.
\eeq
For convenience we will take the relaxation times to be equal: $T^A_2 
= T^B_2 = 1/\Gamma$.

\begin{figure}[t]
%\begin{figure}[!b]
\includegraphics[width=5cm]{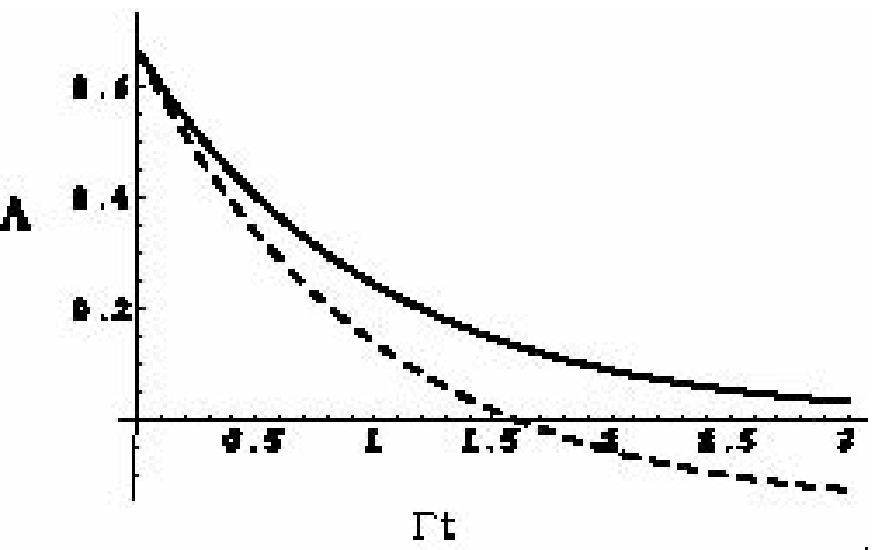}\hspace{0.25in} \includegraphics[width=5cm]{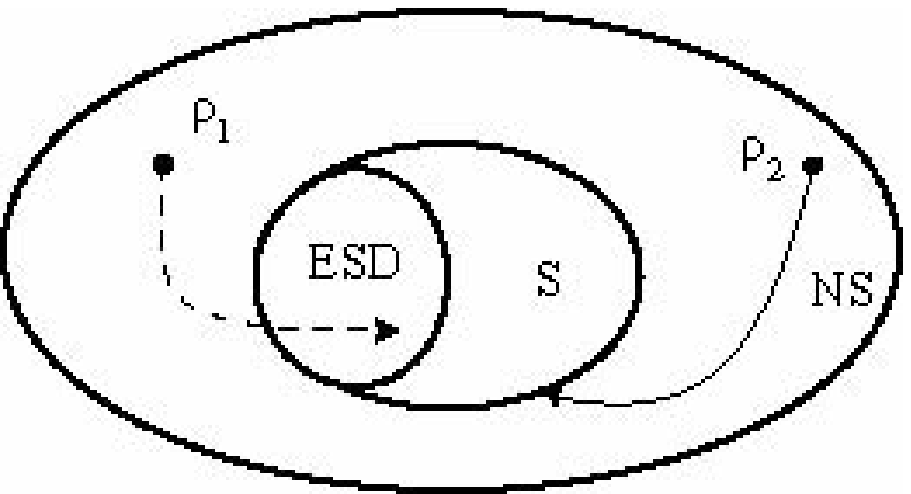}
%%\includegraphics[height=1.5in]{dephasing2n.eps}
%\epsfig{file=FigW-phs.eps, width=6.0 cm}
% Here is how to import EPS art
\caption{{\footnotesize \label{dephas} Two graphs of state trajectories are shown. On the left two $\Lambda$ curves leading to a final separable state, and to the same final value of concurrence, do not reach the same $\Lambda$ value. The ``familiar" asymptotically smooth relaxation of the solid curve, reaching full separability only in the limit $t \to \infty$, is obtained with $a=0, b=c=d=z=1/3$. The dashed curve, obtained for $a=d= 1/12, b=c=z= 5/12$, exhibits ESD by reaching separability at a finite time when $\Lambda$ becomes negative. In the first case $\Lambda$ never becomes negative. The non-metric evolution space picture of the same processes is shown on the right, emphasizing that one trajectory crosses the boundary and the other only asmptotically touches it.}}
\end{figure}

For our purpose it is enough to examine the evolution of $\Lambda$ for an initial state in the more specialized class of bipartite ``X" density matrices 
\cite{YuEberly-arX05}:
\beq \label{mixedrho}
\rho^{AB}(0) =
\begin{pmatrix}
a & 0 & 0 & w\\
0 &  b &  z & 0\\
0 &  z^* &  c & 0\\
w^* & 0 &  0 & d
\end{pmatrix}.
\eeq
where $a+b+c+d = 1$.   The class of mixed X states arises 
naturally in a wide  variety of  physical situations, which include 
pure Bell states as well as the  well-known Werner mixed  states 
\cite{wer}. Here we will take $w = 0$.

From the general solution (\ref{ch}), one can easily show for the 
initial state (\ref{mixedrho}) that one finds for $w=0$
\beq
\Lambda(t) = 2|z(t)|-2\sqrt{ad}.
\eeq
The diagonal elements are independent of $t$, but since $z(t)$ goes to zero as $exp({-\Gamma t})$ as $t$ increases, $\Lambda$ must become strictly negative if $ad \neq 0$. This condition mandates that ESD occurs. The contrasting results that are possible are shown in Fig. \ref{dephas}. One can also check that if both $ad$ and $bc$ are nonzero, then ESD also always occurs for the initial mixed density matrix (\ref{mixedrho}) with both $w$ and $z$ non-zero.

%%%%%%%%%%%%%%%%%%%%%%%%%%%%%%%%%%%%%%%%%%%%
\section{Evolution of $\Lambda$ in Cavity QED}
%%%%%%%%%%%%%%%%%%%%%%%%%%%%%%%%%%%%%%%%%%%%
\label{JC-evolution}

In our second example each of a pair of two-level atoms is confined in a high-Q cavity and exposed to a single mode of radiation resonant with the atomic transition. Each separate atom-mode interaction produces a Jaynes-Cummings evolution governed by separate Hamiltonians \cite{JC}. With $\hbar = 1$, we have 
\beqa
H_{\rm tot} &=& \frac{\om_0}{2}\sigma_z^A  +
\frac{G}{2}(a^{\dagger}\sigma_{-}^A + \sigma_{+}^A a)  + \om a^{\dagger}a \nonumber\\
  && + \frac{\om_0}{2}\sigma_z^B + \frac{G}{2}(b^{\dagger}\sigma_{-}^B +
  \sigma_{+}^B b) + \om b^{\dagger}b
\eeqa
where $\omega_0$ and $\omega$ are the frequencies of atoms and 
cavities, and $G$ is the atom-mode coupling constant, the vacuum Rabi frequency. This model has four interacting parties that we can label in an obvious notation $A, B, a, b$, and it permits well-known exact solutions. 

Here our point is easily illustrated by taking exact resonance and vacuum initial states for the modes and by examining just two sets of entangled initial states for the two atoms  $A$ and $B$, namely superpositions of the two types of Bell states:
\beq
\label{PhiZero1}
|\Phi_{AB}\ra =\cos\alpha|+, +\ra + \sin\alpha|-, -\ra ,
\eeq
and
\beq \label{PZero1}
|\Psi_{AB}\ra = \cos\alpha|+, -\ra + \sin\alpha|-, +\ra .
\eeq

\begin{figure}[!b]
\includegraphics[width=5 cm]{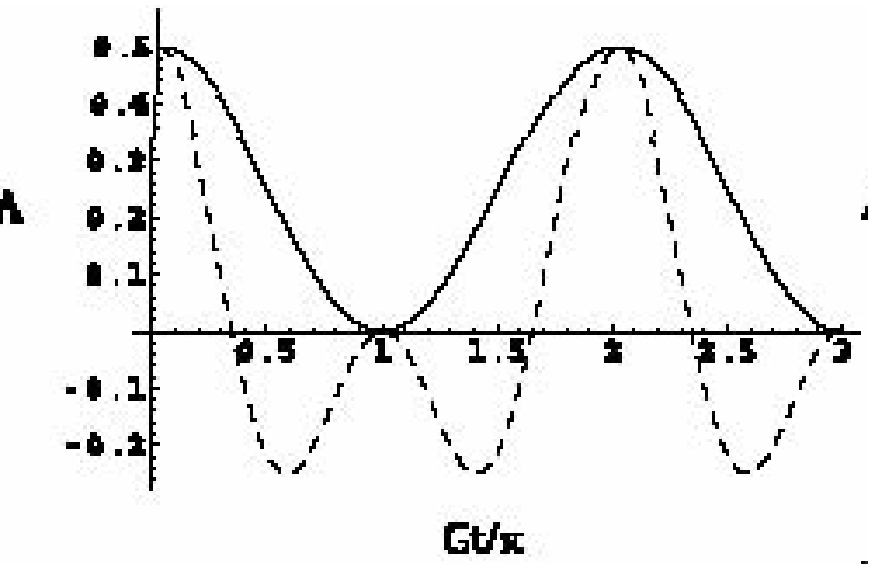}\hspace{0.25in} \includegraphics[width=5cm]{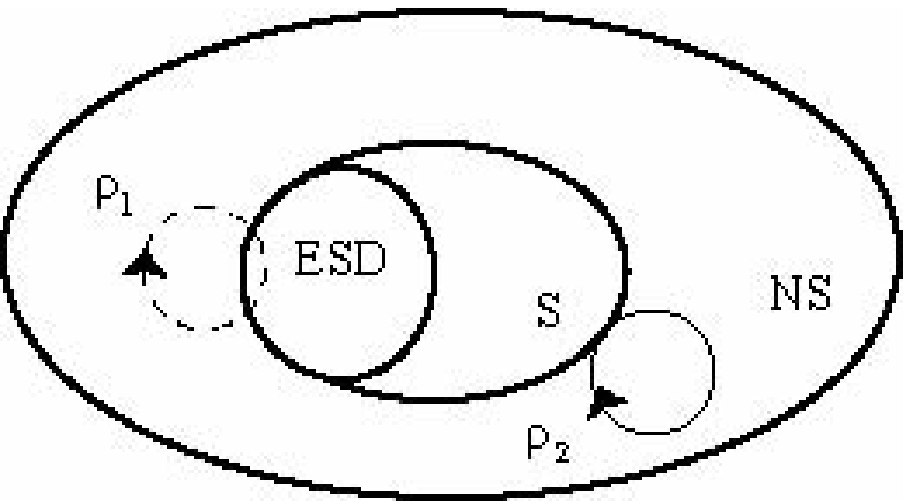}
%\epsfig{file=FigW-phs.eps, width=6.0 cm}
% Here is how to import EPS art
\caption{{\footnotesize \label{JCsoln} Two representations of the same decoherence processes are shown. On the left, two curves of $\Lambda$ vs. $t$ are shown, the first for initial state $\Psi_{AB}$ and the dashed curve for state $\Phi_{AB}$, in both cases for $\alpha = \pi/6$. On the right these periodic processes are seen as periodic trajectories in evolution space, with the interesting feature that one of them repeatedly touches but never crosses the boundary to separability.}}
\end{figure}

A surrogate for ``reservoir" action here is introduced by tracing the radiation modes. This acts as a decoherence process for the $A-B$ pair because it discards completely any information stored in knowledge of the radiation, and as in the preceding example leaves $A-B$ separability as the issue at hand. An earlier examination in which single-mode relaxation was studied \cite{PLKetal02} included an additional physical element by permitting thermal statistics for the mode. In our simpler case, for the initial state (\ref{PhiZero1}) the time-dependent expression for $\Lambda$ turns  out to be \cite{Yonac-Yu-Eberly060c}:
\beq
\Lambda(t) = \cos^2 (Gt/2)\Big(2|\cos\alpha \sin\alpha| - 2\sin^2(Gt/2)|\cos\alpha|^2 \Big).
\eeq
Both of these are shown in Fig. \ref{JCsoln}.

In contrast, for the initial state (\ref{PZero1}), the resulting 
$\Lambda$ is given by
\beq
\Lambda(t)=|\sin2\alpha|\cos^2(Gt/2).
\eeq

%%\begin{figure}[!t]

In the second JC case $\Lambda \geq 0$ for all possible $\alpha$ and $t$, and $\Lambda$ is not zero in any finite time interval (except for the trivial case where the initial state is  separable). We immediately 
conclude that ESD never happens for this type of initial state. However, in the first case the function $\Lambda(t)$ can take negative values. For example, for any $\alpha$ such that $|\cos\alpha| > |\sin\alpha|$,  as $t$ increases from $t=0$ the function $\Lambda(t)$ will take negative values at finite times $t$, and also remain negative  for finite lengths of time. These contrasting behaviors are shown in Fig. \ref{JCsoln}. We see immediately striking differences, compared with Fig. \ref{dephas}, arising from well-known periodic behavior induced by the pure JC interaction.

%%%%%%%%%%%%%%%%%%%%%%%%%%%%%%%%%%%%%%%%%%%%
\section{Concluding comments}
\label{conc}
%%%%%%%%%%%%%%%%%%%%%%%%%%%%%%%%%%%%%%%%%%%%%

In summary,  we undertook to examine the disappearance of entanglment in a finite time (ESD) in relation to considerations suggested by the generic sketch in Fig. \ref{solutionspace} of trajectories of states in ``evolution space". We considered bipartite examples that are explicit, solvable exactly, and simple rather than complicated, to be clear that consequences are not arising from complications but rather from intrinsic properties. We treated only decoherence processes that carry initially entangled states into separable states.

To carry out our discussion we used obvious properties of the separability ``distance" $\Lambda$. We followed two exactly solvable models in which quantum entanglement decreases due to the interaction of qubits $A$ and $B$ with their environments, while they are not interacting with each other, and their environments are also not mutually interacting.  In the second example,  the ``environment" consists of only one cavity mode for each atom, but one can say it acquires the status of environment when it is traced out, i.e., when it is used as an information loss channel. We have pointed out that the $\Lambda$ ``distance" contains information that concurrence does not. In particular, if $\Lambda(\rho)$ becomes negative, it implies that $\Lambda$ describes a separable mixed state.

A few further comments are in order:   (1) The degree of 
entanglement in this paper is measured by the generalized concurrence 
$\Lambda$ which is valid for both pure and mixed states.  For pure 
separable state, one always has $\Lambda=0$. This means that the 
negative $\Lambda$'s only happen for a mixed separable state. (2) 
As usual, we emphasize that entanglement sudden death is independent 
of the choice of measures of  entanglement. (3)  We note that while 
we have found interesting and unexpected  features of $\Lambda$ for 
two solvable models,  it would be very interesting  to  extend these 
results to more general quantum dynamical systems.

A natural question remains for further study: Why does bipartite ESD happen for some initial states, 
but not others, even if both ultimately become separable? The question may be addressed on several 
levels and we believe that a satisfactory answer still does not exist.

%%%%%%%%%%%%%%%%%%%%%%%%%%%%%%%%%%

\section*{Acknowledgments}
%\begin{acknowledgments}

With this small note we add our congratulations at what is probably only the midpoint of the research 
career of Peter Knight. We are among the many admirers of his contributions to quantum information 
and to quantum optics more generally over many years.  Financial support has been received from 
US Army Research Office grant  ARO 48422-PH.
%\end{acknowledgments}

%%%%%%%%%%%%%%%%%%%%%%%%%%%%%%

\bibliography{apssamp}% Produces the bibliography via BibTeX.

%\end{references}

\end{document}